\documentstyle[preprint,aps]{revtex}
\begin{document}
\draft
\title{Molecular Dynamics Simulation of Friction in Xenon Films on a Silver
Substrate}
\author{M. S. Tomassone, J. B. Sokoloff, A. Widom, and J. Krim}
\address{Department of Physics and Center for Interdisciplinary Research
on Complex Systems,\\
Northeastern University, Boston, MA 02115}
\date{\today }
\maketitle

\begin{abstract}
We perform molecular dynamics simulations of friction for atomically thin
Xe films sliding on Ag(111). We determine the inverse of the coefficient
of friction (i.e slip time)  by direct calculation of the 
decay of the center of mass velocity after
applying an external force, as well as from the velocity autocorrelation
function. We find that the slip time
exhibits a  drop followed by a sharp increase in a range of coverage
near one monolayer. The slip time then levels off with further coverage
increases in agreement with previously reported experiments. Our
simulations suggest that the friction found in this system   
is dominated by phonon excitations.
\end{abstract}


\newpage
\narrowtext

Although tribology, the study of friction and wear, has been of
technological interest since ancient times \cite{bowden,dowson}, the
topic continues to rouse interest today \cite
{sokoloff,sokoloff1,widom,daly,dayo,krim1,persson,robbins,persson2,allan}.
Rapid progress in experimental, theoretical and computational methods provides new
insights into the atomic origins of frictional energy dissipation. When a
thin film slides on a metal substrate there exists dissipation of energy via
two mechanisms: (i) electronic excitations in the metallic substrate \cite
{sokoloff1,persson}, and (ii) phonon excitations in the film or in the
substrate \cite{robbins}. The dissipation of energy can 
be characterized by the slip time $\tau$ (i.e. the time it 
takes for the film's speed to fall to $1/e$ of its
original value, assuming it is stopped by friction) or equivalently
by a damping coefficient  $\eta \sim 1/\tau$. 

In this letter we study the phonon contribution 
to friction for Xe films sliding along
a Ag(111) substrate using molecular dynamics simulations. It is of great 
interest to find a way to determine the relative contributions of the phonon 
and electron contributions to friction, since to date it is not clear which 
is  dominant. To this end, we compare our 
results with the slip-time versus coverage data 
reported  by Daly and Krim \cite{daly}, and with the electrical resistivity 
versus coverage data of Dayo and Krim\cite{dayo}. The results of this 
comparison suggest that 
phonon excitations make a dominant contribution to the friction. 

We determine the slip time  as function of 
coverage, defined as the number
of atoms in the film per unit area.
We treat a range of film coverages, from submonolayer to bilayer.
The slip time is determined by two methods. In the first method, an initial
center of mass velocity ${\bf V}_0$ is produced by an external force exerted
on the film for $t<0$. The external force is turned off for $t>0$, and $\tau 
$ is determined by the resulting velocity decay ${\bf V}_0e^{-t/\tau }$ . In the
second method, no external force is applied. The slip time is determined
by the behavior of the  thermal equilibrium autocorrelation  for
the center of mass velocity as a
function of time. The autocorrelation function is related to the 
linear response of the system to a small perturbation by Kubo theory 
\cite{kubo}. Linear response theory 
\cite{allan}  holds in the limit of zero driving velocity and zero 
response force. Therefore, this method
is advantageous for describing
experiments using the Quartz Crystal Microbalance (QCM )\cite{daly,krim1}, 
since the QCM drives the film out of equilibrium
only to a very small extent. This new method for calculating the 
slip-time is  more 
appropriate for the small sliding velocities that occur in the actual 
experiment than previously used methods.  Previous methods  
apply unphysically large 
forces (or, alternatively, shake the substrate at unphysically high frequencies) in order to generate velocities which are several orders of magnitude 
larger than those which occurred in the experiment.

Molecular dynamics simulations of sliding friction have traditionally
employed {\em thermostats} which add or remove energy to or from the system
in order to achieve constant temperature. Thermostats used in
molecular dynamics have the side effect of damping the atomic motions,
resulting in an additional contribution to friction\cite{persson2,robbins}.
In order to study friction in a situation free of such
complications, we employ a thermostat solely to establish statistical thermal
equilibrium, but then turn it off while monitoring the system's 
properties. The absence of a thermostat in the simulation, 
allows us to focus exclusively on the film's dissipation due to phonons.

The model Hamiltonian for $N$ film atoms of mass $m$ 
at positions ${\bf r}_k$ $%
(k=1,..,N)$ is given by 
\begin{equation}
\label{hamilton}H \equiv \sum_{k=1}^{N} \frac{{\bf p}_k^2}{2m}
+U({\bf r}_1,\cdots ,{\bf r}_N), 
\end{equation}
where ${\bf p}_k$ is the momentum of the atom $k$, and the total potential $%
U({\bf r}_1,\cdots ,{\bf r}_N)$ is given by 
\begin{equation}
U({\bf r}_1,\cdots ,{\bf r}_N)\equiv \sum_{k=1}^{N}U_s({\bf r}%
_k)+\sum_{j<k=1}^{ N}V(|{\bf r}_j-{\bf r}_k|). 
\end{equation}
Here, $U_s({\bf r}_k)$ is a single particle potential describing the
interaction between the $k$-th film atom and the substrate, and $V(|{\bf r}%
_j-{\bf r}_k|)$ is the pair potential interaction between the $j$-th and $k$%
-th atoms in the film.

The interaction between two Xe atoms is given by a Lennard-Jones potential
\begin{equation}
V(r)=4\varepsilon \left[ \left( \frac \sigma r\right) ^{12}-\left( \frac 
\sigma r\right) ^6\right] , 
\end{equation}
where $\varepsilon =19.83\,meV$, and $\sigma =4.055\,\AA $ \cite{dash}. The
interaction between a Xe atom and the substrate can be  described
by\cite{steele} a  substrate potential without internal degrees of freedom
 
\begin{equation}
\label{us}U_s({\bf r}_{\Vert },z)=U_0(z)+U_1(z)\sum_{\{{\bf G}\}}\cos ({\bf %
G\cdot r}_{\Vert }), 
\end{equation}
where ${\bf r}_{\Vert }=(x,y)$ are the coordinates of the Xe atom parallel
to the substrate, and $\{{\bf G}\}$ is the set of the six shortest
reciprocal lattice vectors of the substrate. The first term in Eq. (\ref{us}%
) describes the mean interaction of the atoms with the substrate, and the
second term describes the periodic corrugation potential.

Expressions for $U_0(z)$ and $U_1(z)$ were derived by Steele \cite{steele}
assuming that the substrate potential $U_s({\bf r})$ is  a sum of
Lennard-Jones  potentials between one film atom and
all of the atoms in the substrate.  However,
summing Lennard-Jones potentials for $U_s({\bf r})$ is not a correct
description of a metallic surface interaction with a noble gas atom. The
corrugation potential is reduced (from the value found by summing
Lennard-Jones potentials) due to electronic screening. For this 
reason we employ a weaker corrugation potential, 
as did Cieplak {\it et al.} in Ref. \cite{robbins}.
The corrugation potential we use is

\begin{equation}
\label{pote2}U_1(z)=\alpha e^{-g_1z^{*}}\sqrt{\frac \pi {2g_1z^{*}}}\left[ 
\frac{A^{*6}}{30}\left( \frac{g_1}{2z^{*}}\right) ^5-2\left( \frac{g_1}{2z^{*}}%
\right) ^2\right] , 
\end{equation}
where $\alpha =4\pi \varepsilon _{Xe/Ag}A^{*6}/\sqrt{3}$, 
$z^{*}=z/a$, $a=2.892$ $\AA$ 
is the lattice constant of the substrate, $A^{*}=\sigma _{Xe/Ag}/a$, 
$g_1=2/\sqrt{3}$.
We calculate the Lennard-Jones parameters $\sigma _{Xe/Ag} $ 
and $\varepsilon _{Xe/Ag}$ by  fitting 
(i) the position of the minimum of $U_{0}(z)$ 
to the distance between a Xe atom in the first layer and the ion  
cores of the substrate ($z_{0}= 3.6 \AA$ , from \cite{cole}), and 
(ii) the attractive
well depth to the binding energy of one $Xe$ atom to the $Ag$ substrate
$(U_{0}(z_{0})=-211meV$,  from \cite{cole}).
We find  $\sigma _{Xe/Ag}=4.463\,\AA $ and 
$\varepsilon _{Xe/Ag}=13.88\,\,meV$.

Employing Eq. (\ref{pote2}) for the substrate potential
yields a corrugation amplitude of approximately $3\,meV$. This value
is smaller than the corrugation amplitude found using the Steele's
potential ($17.4\,meV$)\cite{steele}. This last corrugation amplitude 
leads to slip times two orders of magnitude smaller than those found in
experiments. 
We have been able to reproduce the results for Kr/Au(111) \cite{krim1,robbins} employing the above approach.

Our simulations are carried out at an equilibrium temperature of 
$T=77.4\,^oK $, and the particles move in a three dimensional box of
size  $ 20\,a \times $ $10\,a\sqrt{3}\times 10\,\sigma$. 
The time scale for vibrations of the adsorbed film
atoms is $t_0=\sqrt{(m\sigma ^2/\varepsilon )}$ $=3.345$ $ps$ , with 
$m=130.1$ $g/mol$. Periodic
boundary conditions in the $x$ and $y$ directions are employed along with a
hard wall boundary condition in the $z$ direction at the top of the box.

We change the coverage by changing the number of Xe atoms $N$.
We use $60\leq N\leq 370$.
All atoms are initially in the gas phase. The atoms condensed in $250\,t_0$\ or
less, forming a triangular lattice incommensurate with the substrate
fcc(111) surface. A thermostat is used only to establish thermal
equilibrium.
To calculate the slip time we use two methods:

{\it Method I}: An external force (parallel to the substrate surface) is
applied to all film particles for approximately $100$ $t_0$. 
The external force then is removed. This induces an
initial center of mass velocity ${\bf V}_0$ in the film which 
decays at later times as $%
{\bf V}_0\,e^{-t/\tau }$. Indeed we find an exponential decay
as shown in Fig. \ref{decay}. 
The slip-time versus coverage results obtained in this manner
are shown in Fig. \ref{sliptime}.
Because the thermostat  was turned off while  the decay of the center
of mass velocity took place, the temperature rose by at most $13\,^oK $
during this process, which occurred for the largest value for the initial 
velocities used in these calculations ($0.6 \, \sigma/t_{0}$).

{\it Method II}: In this method no external force is applied at any time.  
In experiments using  QCM \cite{krim1},
one conventionally describes the data by the 
linear response in the force per unit area $\delta
f$ to an applied 
substrate velocity 
$\delta V$ at complex frequency $\zeta $. 
This response defines the acoustic impedance of the film, $Z(\zeta )\equiv
\lim _{\delta f\rightarrow 0}(\delta f/\delta V)$\thinspace . The Kubo
formula for the acoustic impedance has been derived \cite{allan} to be 
\begin{equation}
\label{kubofla}Z(\zeta )=-i\zeta \rho _2\left( 1+i\zeta \int_0^\infty
C(t)e^{i\zeta t}dt\right) , 
\end{equation}
where $\rho _2\equiv Nm/A$ is the film mass per unit area A, 
and the thermal equilibrium autocorrelation function 
\begin{equation}
\label{corre}
C(t-t^{\prime })\equiv \frac{<V_x(t)V_x(t^{\prime })>}{<V_x(0)^2>}\,\,. 
\end{equation}
In Eq. (\ref{corre}), $V_{x}(t)$ is the center of mass velocity of the 
film in the $x$ direction \cite{comment}. 
The slip time model  \cite{allan} may be written as a Drude-Darcy 
law impedance $Z(\zeta
)=-i(\zeta \rho _2)/(1-i\zeta \tau )$, or equivalently, as the
autocorrelation function  
\begin{equation}
\label{correlation}C(t-t^{\prime })=\exp \left( -\frac{|t-t^{\prime }|}\tau\right). 
\end{equation}

Equation (\ref{correlation}) is the basis of the second method for
calculating $\tau $. The thermal average in Eq. (\ref{corre}) is calculated
performing an average over a time $t_{tot}$

\begin{equation}
\label{average}<V_x(t)V_x(t^{\prime })>=\frac 1{t_{tot}%
}\int_0^{t_{tot}} V_x\left( t+s\right) V_x\left( t^{\prime
}+s\right) ds. 
\end{equation}

In Fig. \ref{sliptime}, the values of the slip time as a function of film
coverage are shown for the two methods described above. The two methods
yield results which are qualitatively similar to each other and to previous
experiments.

We find a minimum followed by a sharp 
rise in the slip time for coverages near one monolayer.
The minimum in $\tau$ corresponds to the uncompressed monolayer
(see Fig. \ref{sliptime}). 
We are able to observe directly in the simulations a compression of the 
monolayer when the coverage varies from $0.0563$ atoms/$\AA^2$ to 
$0.0594$  atoms/$\AA^2$. The interparticle average spacing at
these coverages are respectively: $4.53$ $\AA $
and $4.4$ $\AA $. These values are very close to those reported experimentally
by Unguris et al. \cite{unguris} ( 4.52 $\AA $ and 4.39 $\AA $ for the
uncompressed and fully compressed monolayer).

The structure factor

\begin{equation}
S({\bf Q}) \equiv \frac 1N<\sum_{i,j}^N\cos \left( {\bf Q\cdot }({\bf r}_i-{\bf r}%
_j)\right) > 
\end{equation}
has also been calculated. In Fig. \ref{bragg} we show the Bragg peaks 
in the ${\bf Q}=(Q_x,Q_y,0)$ plane  for different coverages:  
 submonolayer, slightly below uncompressed 
monolayer and compressed monolayer coverages. 
The Bragg peaks for the larger slip times (compressed monolayer)
are sharper
than the Bragg peaks for the smaller slip times (submonolayer and
slightly uncompressed monolayer). Thus, the lattices corresponding to
small slip-times have more
disorder than lattices with large slip-times.

Persson and Nitzan \cite{persson2} have recently calculated friction
for the present system using
molecular dynamics in a model 
that includes a
Langevin thermostat, which they identify with the effect
of electronic excitations in the substrate. 
 This thermostat uses unequal damping constants in directions 
parallel and perpendicular to the sliding direction (i.e.
$\eta _{\parallel }=6.2\times 10^{8}\sec^{-1} $ and 
$\eta _{\bot}=2.5\times 10^{11}\sec^{-1} $.)  
They find a resulting friction constant 
$\eta _{tot }\approx 6.32\times 10^{8}\sec^{-1} $ in their simulations,
for a monolayer coverage.
Assuming  $\eta _{tot}= \eta _{\parallel}+\eta _{phon}$, 
one finds that the phonon contribution to friction 
$\eta _{phon}/ \eta _{total}= 0.02$. Thus there 
appears to be a significant reduction in $\eta _{phon}$ due to
the presence of the electronic damping $\eta _{\parallel}$.
We have performed simulations using the substrate potential
of Eqs. ([4]-[5]) with 
the same Langevin thermostat and damping 
constants as Persson and Nitzan. We find, for a monolayer coverage 
($0.0563$  ${atoms}/\AA^2$ ), that 
$\eta _{total}=2.45\times 10^{9}\sec^{-1}$ and 
$\eta _{phon}=1.83 \times 10^{9}\sec^{-1}$, so that
$\eta _{phon}/ \eta _{total}= 0.75$.    
Thus, although the suppression of the phonon contribution to friction
is significant,
it is not as drastic as that reported in ref. \cite{persson2}.  

In conclusion, we have performed molecular dynamics simulations of
Xe/Ag(111) in a thermostat-free environment.
We have been able to
qualitatively and semi-quantitatively reproduce 
the experimental data \cite{daly} by means of two independent
methods which simulated sliding speeds, which are  
at least three orders of magnitude apart. 
In particular, our simulation reproduces the distinctive 
dip in the submonolayer regime found in experiment. This result, combined 
with the fact that the electrical surface resistivity was found to be nearly 
constant in this regime\cite{dayo} 
suggests that the phonon contribution dominates over the electron 
contribution for this system. The fact that the simulations
yielded comparable slip times for greatly varying sliding speeds
demonstrate the linearity in velocity of the friction
governing this system.

This work has been supported in part by the NSF, Grant No. DMR9204022 and by
the USD of Energy grant No. DE-FG02-96ER45585. We would like to thank Prof.
H. E. Stanley for his aid.  One of us (M. S. T.) would like to thank H.
A. Makse for useful discussions.


\begin{figure}
\caption{Typical decay of 
the center of mass velocity
(in units of $\sigma/t_{0}$) 
after applying the external force.
After the force and thermostat are turned off, 
the velocity decays; 
and this portion of the graph 
is fitted to an exponential curve 
as predicted by the linear friction force law
to obtain the slip time (method I).} 
\label{decay}
\end{figure}

\begin{figure}
\caption{Slip time $\tau$ (in units of sec.) versus coverage (in units of 
number of particles per $\AA^{2}$) for the two methods 
used. (a) Method I. (b) Method II. Solid lines 
in both figures are three of the experimental
curves from Ref. \protect\cite{daly}, included for comparison. There 
is qualitative and quantitative agreement with 
experiment. Both methods reach a minimum near $0.0563$ 
$\mbox{atoms}/\AA^2$.}
\label{sliptime}
\end{figure}

\begin{figure}
\caption{Bragg peaks for (a) a submonolayer coverage, 
$0.040$ $\mbox{atoms}/\AA^2$, (b) slightly 
below uncompressed monolayer coverage,  
$0.0538$ $\mbox{atoms}/\AA^2$, and (c) the compressed monolayer coverage, 
$0.0594$ $\mbox{atoms}/\AA^2$.}
\label{bragg}
\end{figure}


\begin{references}
\bibitem{comment} In our simulations we take $\ V_x$ as
the center of mass velocity minus its mean $<V_{x}>$. 
Subtracting the mean center of mass velocity is 
a standard technique that, in a molecular dynamics simulation, ensures
that the correlation goes to zero at long times \protect\cite{primer}.
\end{references}
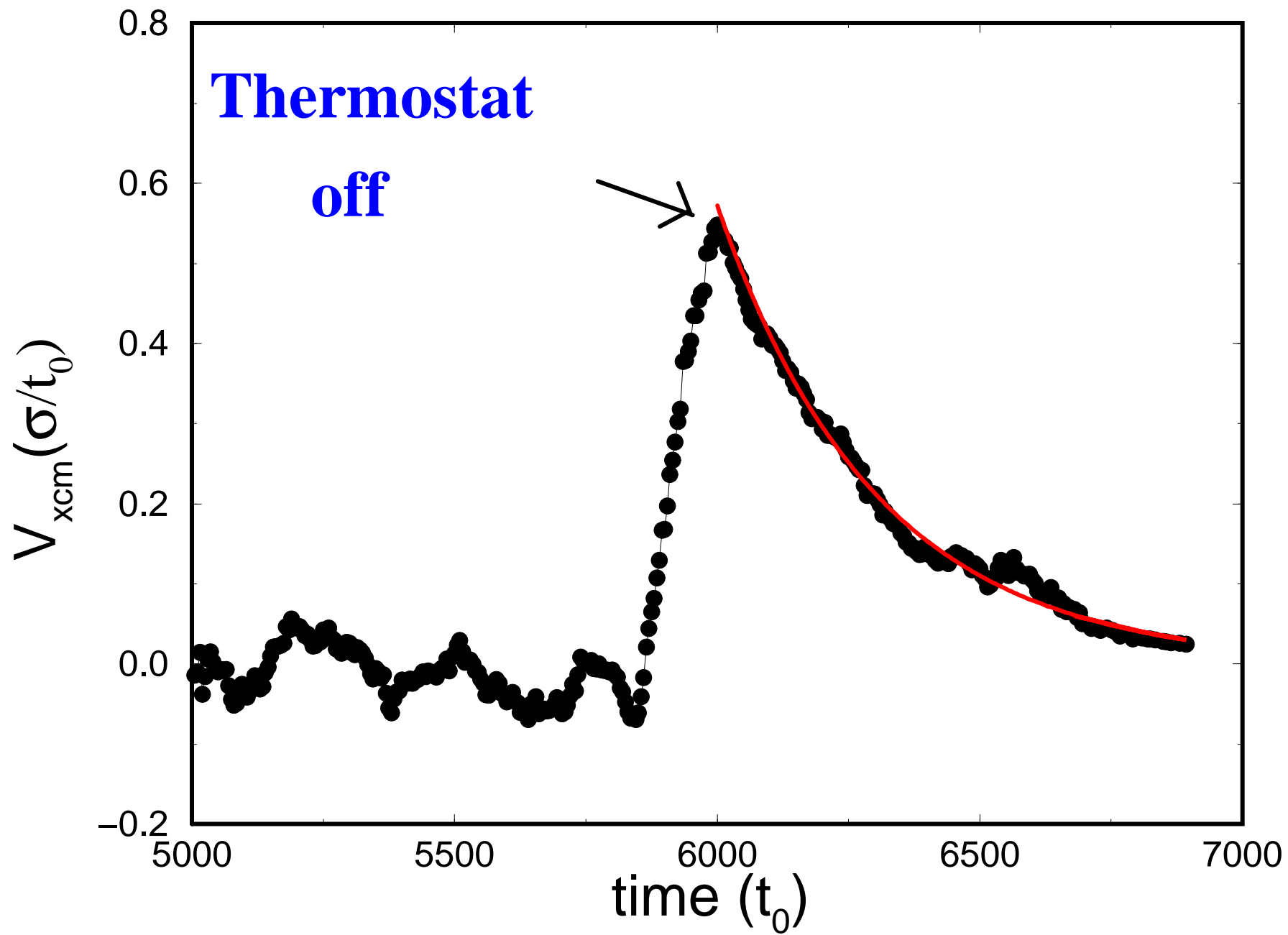
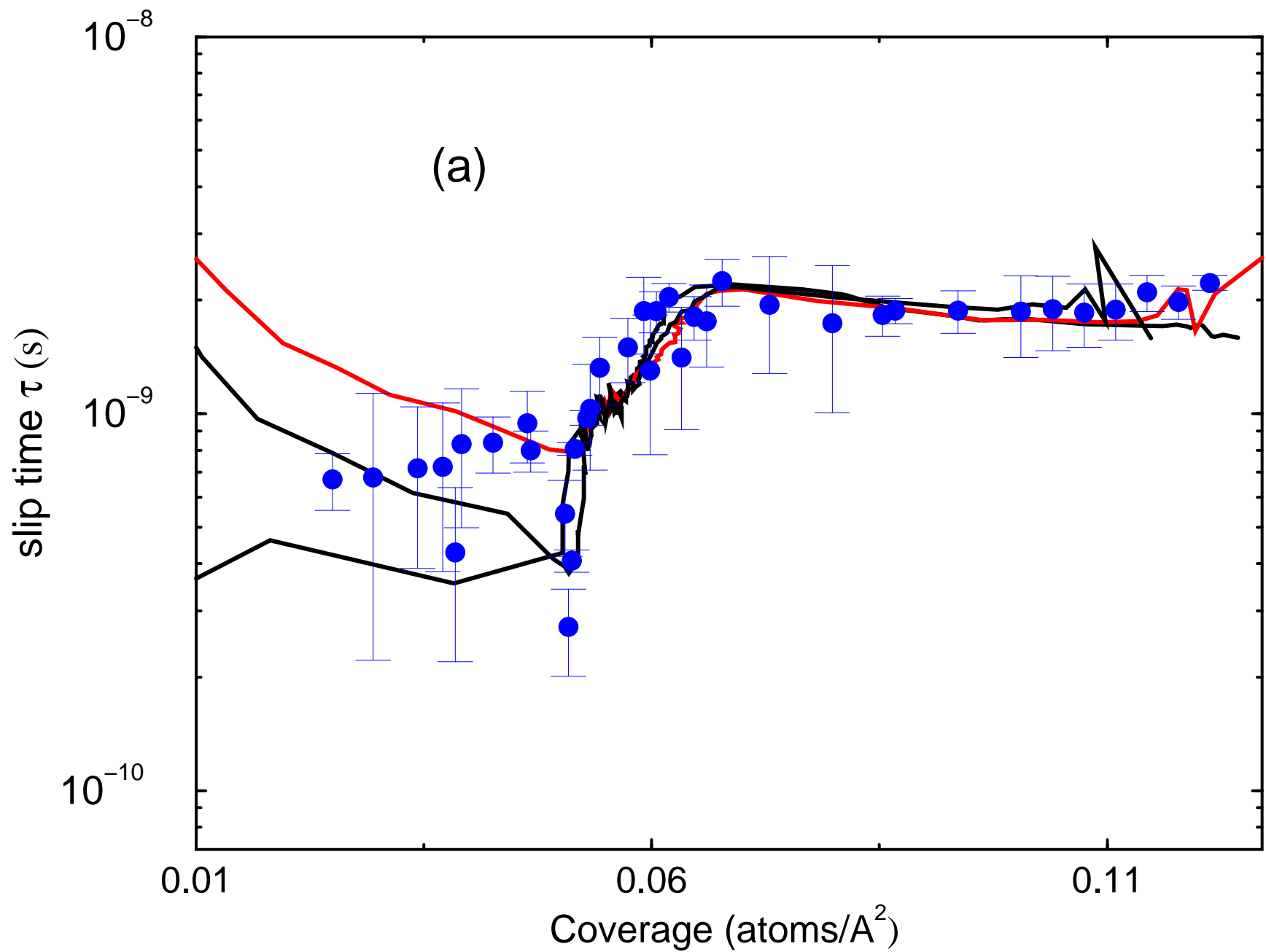
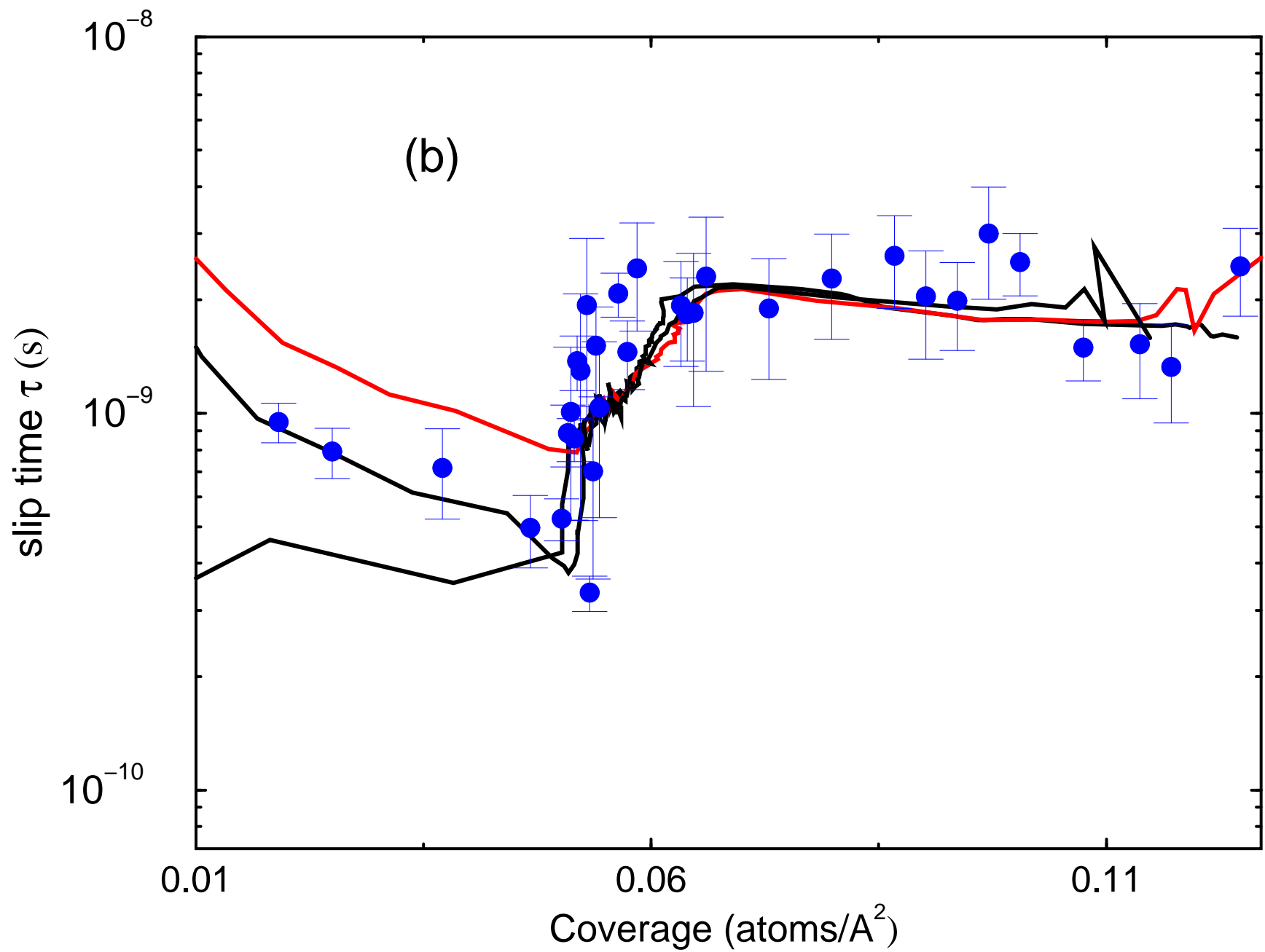
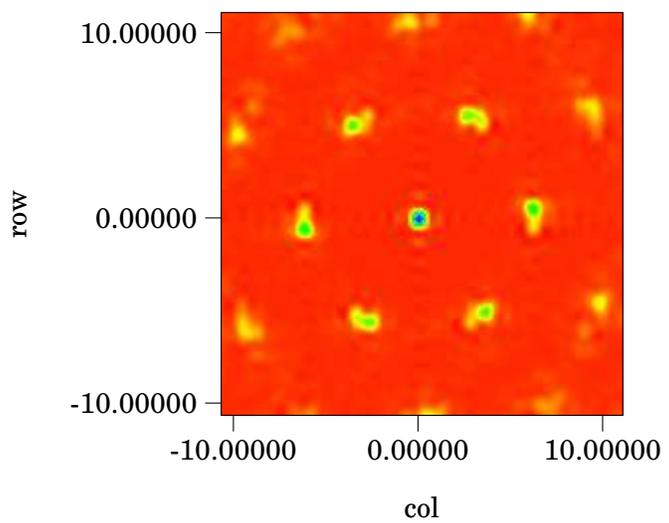
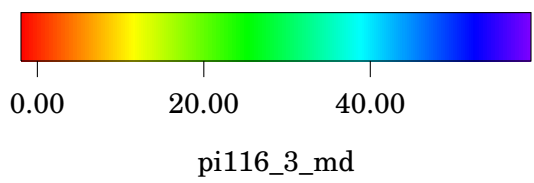
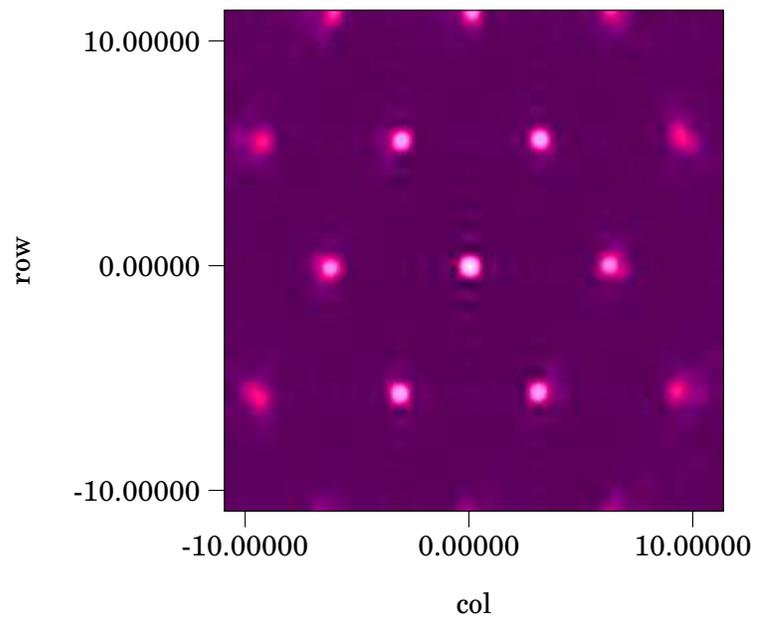
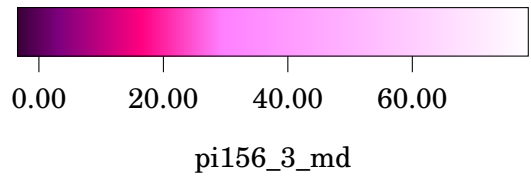
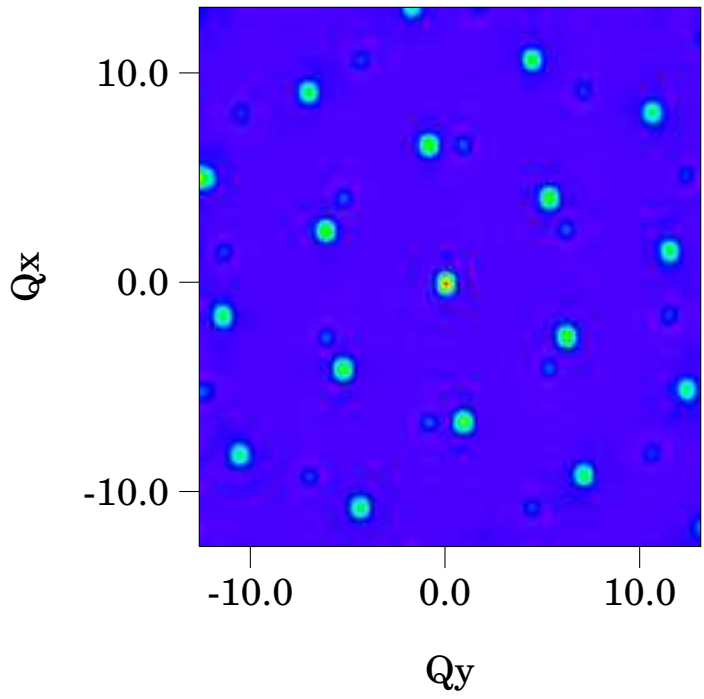
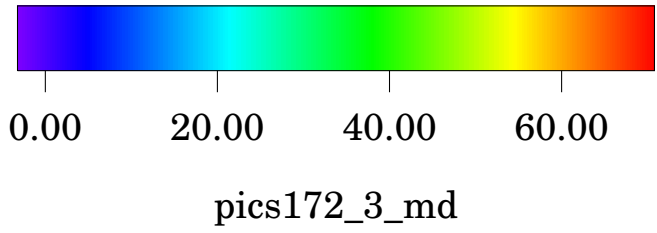
\end{document}